\documentclass[10pt, conference, letterpaper]{IEEEtran}
\IEEEoverridecommandlockouts
\usepackage{cite}
\usepackage{amsmath,amssymb,amsfonts}
\usepackage{algorithmic}
\usepackage{graphicx}
\usepackage{textcomp}
\usepackage{xcolor}
\usepackage{url}
\usepackage{physics}
\usepackage{cuted}
\usepackage{subcaption}
\def\BibTeX{{\rm B\kern-.05em{\sc i\kern-.025em b}\kern-.08em
    T\kern-.1667em\lower.7ex\hbox{E}\kern-.125emX}}
\begin{document}

\title{Benchmarking Classical and Quantum Optimization Approaches for Rider–Order Assignment
}

\author{\IEEEauthorblockN{Tharrmashastha SAPV}
\IEEEauthorblockA{\textit{IIIT-Delhi, India} \\
\textit{EulerQ Softwares pvt Ltd., India}\\
tharrma@eulerq.com}
\and
\IEEEauthorblockN{Surya Prakash Palanivel}
\IEEEauthorblockA{\textit{EulerQ Software pvt Ltd., India} \\
surya@eulerq.com}
\and
\IEEEauthorblockN{Jasjyot Singh Gulati}
\IEEEauthorblockA{\textit{IIIT-D, India} \\
jasjyot23257@iiitd.ac.in}
\and
\IEEEauthorblockN{M Maruthu Pandi}
\IEEEauthorblockA{\textit{EulerQ Software pvt Ltd., India} \\
maruthu@eulerq.com}
}

\maketitle

\begin{abstract}
The logistics industry is widely regarded as a promising application domain for emerging optimization paradigms, including quantum computing. The Rider–Order Assignment problem is a practically motivated optimization problem arising in online food delivery and related logistics applications. While the problem is closely related to the classical matching problem, the inclusion of realistic operational constraints renders it computationally challenging.
In this work, we formulate the Rider-Order Assignment problem as a constrained binary optimization problem and perform a comparative analysis of classical, quantum-inspired, and gate-based quantum solvers for this problem across multiple instance sizes. Solver performance is assessed using solution quality, computational runtime, and constraint satisfaction, with a consistent post-processing procedure applied to ensure feasibility.
\end{abstract}

% \begin{IEEEkeywords}
% quantum computing,
% optimization,
% logistics,
% quantum algorithm
% \end{IEEEkeywords}

Quantum computing is an emerging computational paradigm that exploits the principles of quantum mechanics to perform computations and has demonstrated potential speedups over classical approaches for certain computational problems. Its applications are expected to impact multiple industries, including logistics(\cite{log1, log2}), pharmaceuticals(\cite{pharma1, pharma2}), automotive(\cite{automotive1, automotive2}), and banking(\cite{banking1, banking2}). Within the logistics domain, online food delivery represents a particularly optimization-intensive sector. A recent report~\cite{globalreport} estimates that the global online food delivery services market was valued at \$380.43 billion in 2024 and is projected to grow to \$618.36 billion by 2030, a trend driven by increasing technology penetration worldwide.

At scale, food delivery platforms process an enormous volume of orders: for example, DoorDash handles approximately $7$ million orders per day on average~\cite{doordashdelivery}, while in rapidly growing markets such as India, platforms like Swiggy and Zomato process between 1 and 2 million orders daily. One of the most significant operational challenges in these systems is accurately assigning delivery riders to incoming orders. We refer to this problem as the Rider–Order Assignment problem. At the time an order is placed, multiple delivery partners may be available. Assignments must be made in a manner that minimizes operational costs, while maximizing rider utilization, and satisfying multiple operational constraints.

At its simplest, the rider–order assignment problem can be viewed as a variant of the classical matching problem. The problem may be represented as a bipartite graph with orders and riders forming the two partitions, where each potential rider–order pair is associated with a preference score that reflects operational costs such as travel distance, expected delivery time, and related factors. The goal is to determine a matching that satisfies a set of operational constraints.

Under simplified constraints, such as requiring each order to be assigned to exactly one rider and vice versa, the problem can be solved in polynomial time using standard matching algorithms. However, in real-world operations, such constraints are rarely sufficient. Practical constraints commonly include allowing multiple orders per rider, imposing upper bounds on rider capacity, and enforcing service-level agreements. The introduction of these constraints significantly increases the complexity of the rider–order assignment problem and renders it intractable for existing polynomial-time matching algorithms. In particular, imposing capacity constraints on riders and limiting the number of orders per rider to a constant other than one is sufficient to push the problem into the NP-hard regime. 
As a result, exact optimization approaches become impractical at scale, and one must instead rely on approximate solvers that operate in polynomial time to produce near-optimal solutions, though not necessarily optimal. For optimization problems of this nature, quantum algorithms may offer a promising alternative by providing near-optimal solutions within practical computational budgets.

% Although practical fault-tolerant quantum computers are a few years away, there has been significant research on how near-term and early fault-tolerant quantum hardware can be effectively leveraged. This has led to the development of Variational Quantum Algorithms (VQAs), a class of hybrid classical–quantum algorithms that combine low-depth parametric quantum circuits for energy estimation with classical optimization routines for tuning circuit parameters. A prominent subclass of VQAs is the Quantum Approximate Optimization Algorithm (QAOA), a gate-based variational algorithm inspired by continuous-time quantum annealing and specifically designed to address hard combinatorial optimization problems. While QAOA is a heuristic method and does not guarantee a quantum advantage, it provides a practical framework for approximately solving challenging optimization problems using shallow quantum circuits.

In this work, we formally present the Rider-Order Assignment problem as a constrained binary optimization problem. This formulation captures the key operational costs and constraints encountered in real-world food delivery services.
Using this formulation, we perform a comparative study across multiple solver paradigms, including classical exact and heuristic methods, quantum-inspired Ising-based solvers, and gate-based variational quantum algorithms. We mainly focus on the trade-offs among solution quality, computational runtime, and constraint satisfaction across different problem scales. 
Rather than claiming quantum advantage, this study aims to provide a careful assessment of the strengths, limitations, and practical applicability of these solver approaches for large-scale, constrained assignment problems.

\section{Formulation of the Rider-Order Assignment Problem}
\label{sec:formulation}

We now present the formulation of the Rider–Order Assignment problem considered in this work. Orders are processed in batches over a fixed time interval, and each batch consists of $n$ orders and $m$ available riders. The objective of the assignment problem is to assign orders to riders so as to minimize a cost function defined over each batch.

The objective function consists of four components: (i) the distance between a rider and the corresponding pickup location, (ii) the rider’s travel time to the pickup location, (iii) rider or restaurant wait time, and (iv) a fairness term that encourages balanced order allocation across riders. The optimization is performed subject to a set of hard and soft constraints.

The hard constraints are as follows:
\begin{itemize}
    \item[(a)] Order Assignment constraint, which enforces that each order is assigned to exactly one rider,
    \item[(b)] Rider Load constraint, which limits the number of orders assigned to a rider to at most a fixed constant $k$, and
    \item[(c)] Rider Capacity constraint, which upper bounds the total size of the orders assigned to a rider.
\end{itemize}
In addition, we impose two soft constraints: 
\begin{itemize}
    \item[(d)] Rider Geo-fencing constraint, that restricts the pickup distance to lie within a fixed radius, and
    \item[(e)] Time Service-Level Agreement (SLA) constraint, that upper bounds the allowed time till delivery since the time of order placement.
\end{itemize}
Since constraints (d) and (e) may not be satisfiable for all batches in practice, we model them as soft constraints.
When feasible, these constraints are enforced directly. But when no feasible solution exists, violations are penalized in the objective function.

Mathematically, we define the Rider Assignment problem as follows.

\begin{align*}
    \min &~~\alpha \sum_{i,j} RD^p_{i,j}\cdot x_{i,j} + \beta \sum_{i,j} OT^d_{i,j}\cdot x_{i,j} \\
    &~~~~~~~~~~~~+ \gamma \sum_{i,j} WT_{i,j}\cdot x_{i,j} + \delta \sum_i \Big[CO_i + \sum_j x_{i,j}\Big]^2\\
    \text{s.t.} &~~\bullet\sum_{i} x_{i,j} = 1~~\forall j\in n\\
    &~~\bullet \sum_{j} x_{i,j} \le k~~\forall i \in m\\
    &~~\bullet \sum_{j}S_j\cdot x_{i,j} \le SC_i~~\forall i\in m\\
    &~~\bullet RD^p_{i,j}\cdot x_{i,j} \le GF ~~\forall~i\in m, j\in n\\
    &~~\bullet \Big[\max\big\{RT^p_{i,j},~PT_j\big\} + RT^d_{i,j}\Big]\cdot x_{i,j} \le PR_j ~~\forall~i,j
\end{align*}
where $x_{i,j} \in \{0,1\}$ is the indicator variable that indicates of rider $i$ is assigned order $j$, $RD^p_{i,j}$ is the distance between rider $i$ and pickup location of order $j$, $OT^d_{i,j}$ is the time taken by rider $i$ to deliver order $j$ after pickup, $WT_{i,j}$ is the time rider $i$ waits in the restaurant to pickup order $j$ or the time restaurant waits before rider $i$ picks up order $j$, $CO_i$ is the number of orders delivered by rider $i$ up to the current batch, $SC_i$ is the size capacity of rider $i$, $GF$ is an upper bound on the preferred pickup distance, $PT_j$ is the prepare time of order $j$, $RT^p_{i,j}$ is the time taken for rider $i$ to reach pickup location of order $j$, $RT^d_{i,j}$ is the time taken for rider $i$ to deliver order $j$, and $PR_j$ is the duration within which the delivery is promised.

The objective function is designed to capture key operational considerations in rider–order assignment. The first term penalizes the distance between a rider and the pickup location, encouraging geographically efficient assignments and reducing unnecessary rider travel. The second term minimizes the time a rider takes to deliver an order after pickup, accounting for service efficiency and delivery responsiveness.
The waiting-time term captures both the time a rider spends waiting at a restaurant for an order to be prepared and the time a restaurant waits for a rider to arrive for pickup. Minimizing this quantity improves rider utilization while simultaneously ensuring timely pickups and preserving order freshness. Finally, the fairness term promotes balanced workload distribution among riders by penalizing skewed order assignments. Such imbalances can lead to reduced rider satisfaction and increased attrition, which in turn results in higher operational costs for delivery platforms.

Similarly, the constraints in the formulation are designed to reflect operational requirements encountered in real-world delivery systems. The order assignment constraint enforces that each order is assigned to exactly one rider, while the rider assignment constraint limits the maximum number of orders that can be allocated to a single rider. In addition, each rider has a finite carrying capacity, and the rider capacity constraint ensures that the total volume of assigned orders does not exceed this limit.
The geofencing constraint captures the practical preference of riders to accept nearby pickup locations, thereby reducing the likelihood of order reassignments due to excessive pickup distances. The service-level agreement (SLA) constraint enforces timely delivery by restricting the allowable delivery time for each order. Violations of the assignment, load, and capacity constraints can lead to operational inconsistencies and are therefore modeled as hard constraints. In contrast, the geofencing and SLA constraints are preferential in nature and are modeled as soft constraints, as satisfying them for all orders may not always be feasible in practice.

An important modeling assumption in this formulation is that order costs are additive, meaning that the cost of assigning multiple orders to a rider is approximated as the sum of individual order costs. In practice, the true delivery cost depends on routing and sequencing decisions across pickup and drop-off locations, which introduces additional complexity. In this work, we adopt a simplified first-stage assignment model, where the additive cost serves as a good proxy for the true delivery cost and often provides an upper bound on the realized cost.

\section{Experimental Setup}

\subsection{Solvers}
\label{sec:solvers}

In this work, we use the following solvers to solve the Rider-Order Assignment problem:
\begin{itemize}
    \item[(i)] \textit{Greedy}: The Greedy solver serves as a simple heuristic baseline for the rider-order assignment problem. In this approach, orders are processed sequentially, and each order is assigned to the rider that minimizes a local criterion, namely the distance to the pickup location as long as the size of the order is within the rider capacity. Although distance-based selection is used in this work, alternative greedy criteria such as minimizing delivery time or prioritizing riders with fewer assigned orders are also possible. This solver does not explicitly account for global optimality or constraint interactions and is included primarily as a lightweight baseline.

    \item[(ii)] \textit{SCIP}: \textit{SCIP} is an open-source solver for constraint integer programming and mixed-integer optimization. It solves the rider-order assignment problem as a mixed-integer quadratic program, handling binary decision variables and linear constraints natively while incorporating quadratic terms directly in the objective. SCIP employs a combination of branch-and-bound, LP relaxations, cutting planes, and constraint propagation to search for optimal or near-optimal solutions. In this study, SCIP is used as a classical exact solver benchmark against which heuristic and quantum-inspired solvers are compared. We use the SCIP interface provided through the \textit{JijModeling} Python package~\cite{jijmodeling}.
    
    \item[(iii)] \textit{Simulated Quantum Annealing (SQA)}: Simulated Quantum Annealing is a Markov Chain Monte Carlo (MCMC)–based heuristic that approximates the dynamics of quantum annealing using classical computation. By emulating quantum tunneling effects, SQA can, in certain regimes, more effectively explore optimization landscapes containing narrow, high-energy barriers than purely classical local-search methods. As a result, SQA has been studied as a quantum-inspired approach for solving combinatorial optimization problems formulated in QUBO or Ising form. In this work, we use the SQA implementation provided by \textit{OpenJij}\cite{OpenJij}.

    \item[(iv)] \textit{Coherent Ising Machine (CIM)}: The Coherent Ising Machine is a quantum-inspired hardware approach for solving combinatorial optimization problems formulated as Ising or QUBO models. A CIM is implemented as a network of coupled optical parametric oscillators, where binary spin variables are encoded in the phase states of optical pulses and interactions between spins are realized through optical coupling. 
    % The system evolves toward low-energy configurations through a combination of gain saturation, dissipation, and nonlinear dynamics, effectively performing an analog optimization process. 
    While CIMs do not constitute universal quantum computers, they exploit physical processes inspired by quantum dynamics and have been studied as a promising platform for large-scale Ising optimization. In this work, we use CIM hardware of Quanfluence, which provides cloud-based execution and readout of Ising optimization instances.
    
    \item[(v)] \textit{QAOA}: The Quantum Approximate Optimization Algorithm(\cite{qaoa}) (QAOA) is a gate-based variational quantum algorithm inspired by continuous-time quantum annealing. It follows a hybrid quantum–classical workflow in which a parameterized quantum circuit alternates between problem and mixer Hamiltonians for a fixed number of layers $p$, while a classical optimizer updates the circuit parameters. Although increasing $p$ improves expressivity, it also leads to circuits or high depth that are difficult to simulate in the near-term regime. Consequently, we restrict our study to shallow circuits with $p=3$ and perform classical parameter optimization using the COBYLA optimizer. In this work, QAOA is used as a penalty-based variational approach, with constraints incorporated through the QUBO formulation rather than enforced at the circuit level.

    \item[(vi)] \textit{QAOAnsatz}: The Quantum Alternating Operator Ansatz (QAOAnsatz) is an extension of the standard Quantum Approximate Optimization Algorithm designed to address constrained optimization problems by restricting the quantum evolution to the feasible subspace. Instead of employing the standard transverse-field mixer, QAOAnsatz uses problem-specific mixer Hamiltonians that preserve constraint satisfaction throughout the evolution. Early analytical work~\cite{xymixer} demonstrated that replacing the transverse-field mixer with an $XY$-type mixer allows one-hot constraints to be preserved, thereby preventing transitions to invalid configurations, and subsequent studies~\cite{fuchsxy} generalized this idea to arbitrary feasibility conditions, ensuring that the quantum state evolves only within the space of valid solutions. In this work, the QAOA Ansatz is used to explicitly enforce the order assignment constraint via the mixer Hamiltonian, while the other constraints are enforced via penalties.
\end{itemize}

\subsection{QUBO Formulation of the Rider-Order Assignment Problem}

The Rider–Order Assignment problem formulated in Section~\ref{sec:formulation} is a constrained binary optimization problem. Among the solvers considered in this work, all except SCIP and the Greedy heuristic operate on Quadratic Unconstrained Binary Optimization (QUBO) formulations. Consequently, to apply these solvers, the constrained problem must be transformed into an equivalent QUBO representation.

A standard approach for converting optimization problems with quadratic objectives and linear constraints into QUBO form is the penalty method. In this approach, each constraint is incorporated into the objective function through an additional penalty term weighted by a corresponding penalty coefficient. Therefore, constraint violations increase the objective value and bias the optimization away feasible solutions. When penalty coefficients are chosen appropriately, feasible solutions are favored while preserving the structure of the original objective.

In the QUBO formulation, the order assignment constraint is enforced by converting it into an equality constraint of the form $g(x) = 0$, and the corresponding penalty term $(g(x))^2$ added to the objective function. Similarly, the rider load and rider capacity constraints are originally expressed as inequality constraints of the form $g(x) \leq 0$. These constraints are converted into equality constraints by introducing non-negative slack variables $s$, yielding $g(x) + s = 0$, and are penalized using quadratic terms of the form $(g(x) + s)^2$.

The soft constraints are handled differently. Although these constraints apply to individual assignment variables $x_{i,j}$, directly enforcing them through preprocessing, by fixing variables that violate the constraints, can eliminate high-quality solutions or render the optimization problem infeasible, since these constraints are preferential rather than mandatory. Instead, each soft constraint of the form $a_{i,j} x_{i,j} \leq b_{i,j}$ is incorporated into the objective through a penalty term of the form $\max\{(a_{i,j} - b_{i,j}), 0\} \cdot x_{i,j}$ that assumes a positive value upon violation, while no penalty is incurred otherwise.

Incorporating these, the QUBO corresponding to the Rider-Order Assignment problem is:
\vspace{-0.6cm}
\begin{strip}
\begin{align*}
    \min & ~~\alpha \sum_{i,j} RD^p_{i,j}\cdot x_{i,j} + \beta \sum_{i,j} OT^d_{i,j}\cdot x_{i,j} + \gamma \sum_{i,j} WT_{i,j}\cdot x_{i,j} + \delta \sum_i \Big[CO_i + \sum_j x_{i,j}\Big]^2 + \lambda_A \cdot \sum_j \Big( \sum_i x_{i,j} - 1\Big)^2\\
    &~~~~~~~~~~~~+ \lambda_L \cdot \sum_i \Big( \sum_j x_{i,j} +r_{i}-k\Big)^2 + \lambda_{Cap} \cdot \sum_i \Big( \sum_j S_j\cdot x_{i,j} + s_{i} - SC_i\Big)^2\\
    &~~~~~~~~~~~~+ \lambda_{GF} \cdot \sum_i \sum_j \max\Big\{ RD^p_{i,j} -GF, 0\Big\} \cdot x_{i,j} + \lambda_{P} \cdot \sum_i \sum_j \max\Big\{ P^{coeff}_{i,j} - PR_j, 0\Big\} \cdot x_{i,j}\\
    \text{s.t.} &~~~x_{i,j}\in \{0,1\} ~~\forall~i,j
\end{align*}
\end{strip}
where $P^{coeff}_{i,j} = \max\big\{RT^p_{i,j},~PT_j\big\} + RT^d_{i,j}$. The values of the penalty multipliers $\lambda$ have been chosen in an instance-dependent manner.

\subsection{Problem Instances}
\label{sec:instances}

We define a problem instance as a set of $m$ riders and $n$ orders, and refer to it as an $(m,n)$ instance. In this work, we consider instances with $m = n$, and define the size of an instance to be $n$. The maximum number of orders that can be assigned to a rider is fixed to $k = 2$ across all experiments.

Due to the quadratic scaling of the number of binary decision variables with instance size, different solver classes are evaluated in different problem-size regimes. QAOA and QAOAnsatz, are evaluated on small instances with $n \le 4$, since classical simulation of large quantum circuits remains hard. In contrast, Simulated Quantum Annealing (SQA) and the Coherent Ising Machine (CIM) are evaluated on larger instances with $n \ge 10$. Within each size regime, all solvers are evaluated on identical problem instances to ensure fair comparison.

For each instance size, we generate 40 independent problem instances modeling delivery operations within the city of Bangalore. Rider and order locations are sampled uniformly at random within the city limits, and attributes, such as rider capacity, completed orders, order size, preparation time, and promised delivery time, are sampled from realistic operational ranges. Distances and travel times are computed using map-based routing from OpenStreetMap, augmented with a traffic model trained on real-time Bangalore traffic data. To ensure reproducibility, random seeds are fixed per instance size, and objective function coefficients are normalized prior to optimization.

\subsection{Evaluation Metrics}

We evaluate all solvers using three primary metrics: objective value, runtime, and constraint satisfaction. The objective value corresponds to the total cost of the optimization problem, including penalty contributions arising from constraint violations in the QUBO formulation. For SCIP, hard constraints are enforced natively and therefore incur no penalty contributions. A similar behavior is observed for the Greedy heuristic, which produces feasible assignments by construction. In contrast, penalty-based solvers may return solutions that violate one or more hard constraints.

To ensure comparability across solvers, we apply a post-processing repair procedure to all solutions that violate hard constraints. This procedure resolves infeasible assignments by addressing orders assigned to multiple riders or to no rider, as well as riders exceeding the maximum allowed number of assigned orders. We refer to such cases as hanging orders and hanging riders, respectively. Whenever possible, mutually preferable rider--order assignments are fixed first. Remaining infeasible assignments are resolved by solving a reduced instance of the assignment problem restricted to the hanging riders and orders. Unless otherwise stated, objective values reported for comparison correspond to the post-repair solutions.

Runtime is measured as the average wall-clock time required by each solver to process a single problem instance. Due to the wide variation in computational cost across solver paradigms, runtime results are reported on a logarithmic scale.

Constraint satisfaction is reported in terms of the number of violated soft constraints in the final solution. This metric quantifies the extent to which preferential constraints, such as geofencing and delivery time SLAs, are satisfied after feasibility repair.

\section{Results}

In this section, we present a comparative evaluation of the solvers introduced in Section~\ref{sec:solvers} on the rider--order assignment problem. All results are averaged over 40 randomly generated instances per problem size, as described in Section~\ref{sec:instances}.  
Unless otherwise stated, solution quality is evaluated using the post-repair objective value, ensuring feasibility with respect to hard constraints. We report results separately for small and large instance regimes, reflecting the computational feasibility of the different solver classes.

\subsection*{Solution Quality}

\begin{figure}[h]
    \centering
    \includegraphics[width=0.95\linewidth]{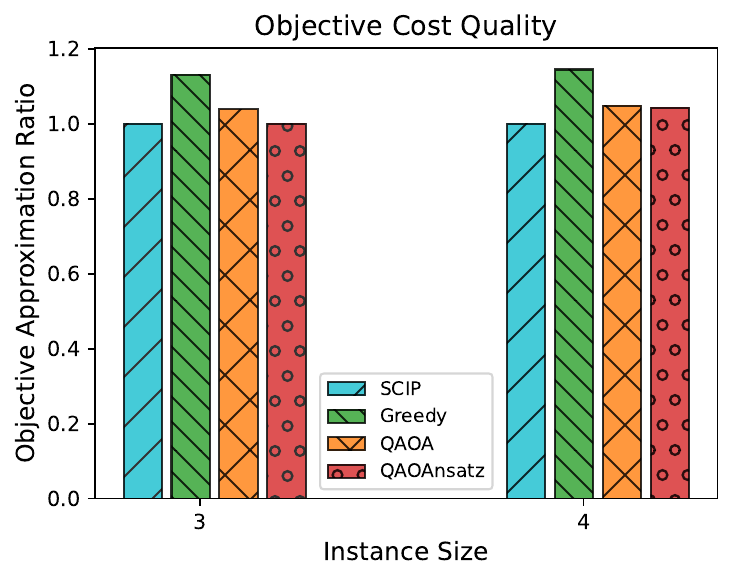}
    \caption{The Average objective cost of the small instance solvers normalized by the objective cost of SCIP.}
    \label{fig:small-obj}
\end{figure}

\begin{figure}
    \centering
    \includegraphics[width=\linewidth]{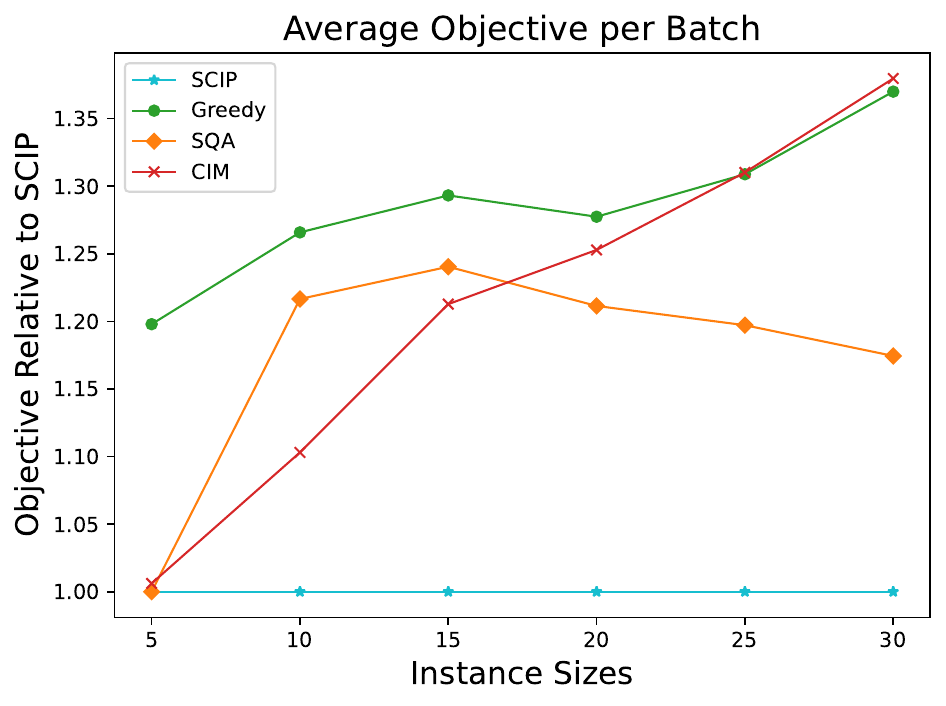}
    \caption{The Average objective cost of the large instance solvers normalized by the objective cost of SCIP.}
    \label{fig:large-obj}
\end{figure}

Across both small and large instance regimes, classical solvers provide a strong baseline in terms of post-repair objective value. For small instances~(Fig.\ref{fig:small-obj}), both QAOA and QAOAnsatz achieve objective values that are competitive with SCIP. Among quantum approaches, the QAOA Ansatz typically produces lower objective values than standard QAOA, reflecting the benefit of explicitly enforcing a single hard constraint at the mixer level rather than relying solely on penalty terms.

For larger instances~(Fig.\ref{fig:large-obj}), quantum-inspired solvers exhibit mixed behavior. SQA produces feasible solutions with objective values that are moderately higher than those obtained by SCIP, while CIM demonstrates slightly worse solution quality than SQA for certain instance sizes. Although the performance of the quantum-inspired solvers seems poor, their ability to achieve similar quality even with a very large number of variables in less time makes them suitable for real-world optimization problems.

\subsection*{Runtime Performance}

\begin{figure}
    \centering
    \includegraphics[width=1\linewidth]{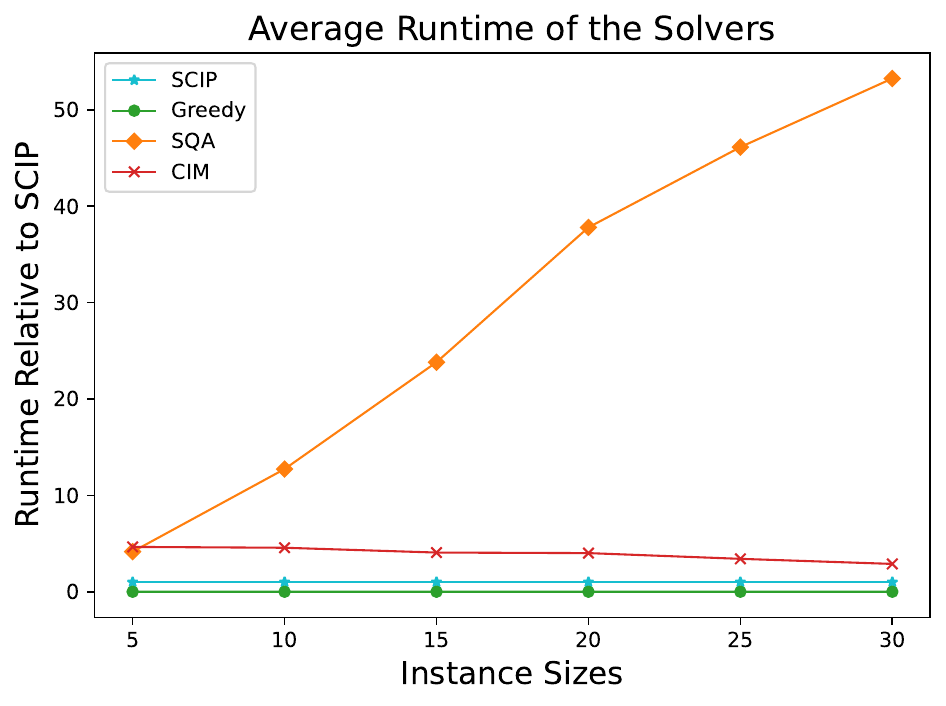}
    \caption{The Average runtime of the large instance solvers normalized by the runtime of SCIP.}
    \label{fig:large-runtime}
\end{figure}

Runtime measurements reveal significant variation across solver paradigms~(Fig.\ref{fig:large-runtime}). As expected, the Greedy solver consistently exhibits the lowest runtimes across all instance sizes considered. SCIP follows the greedy solver. CIM incurs slightly higher computational time but decreases with increasing instance size. In contrast, SQA incurs substantially higher computational costs, with runtime increasing rapidly as the instance size grows. This behavior in SQA is possibly a consequence of the slow cooling times and the exponential increase in parameter space as instance size increases.

\subsection*{Constraints Violation}

\begin{figure}
    \centering
    \includegraphics[width=\linewidth]{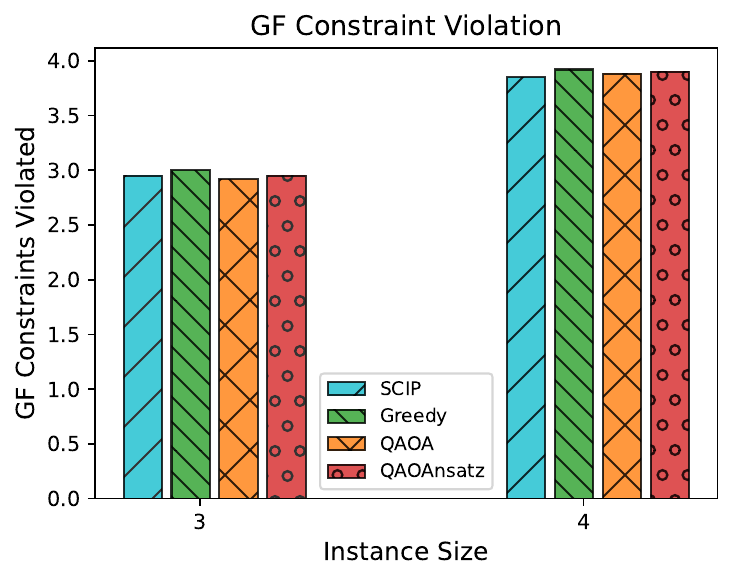}
    \caption{The GF Constraint violations for small instance solvers.}
    \label{fig:small-gf}
\end{figure}

\begin{figure}
    \centering
    \includegraphics[width=\linewidth]{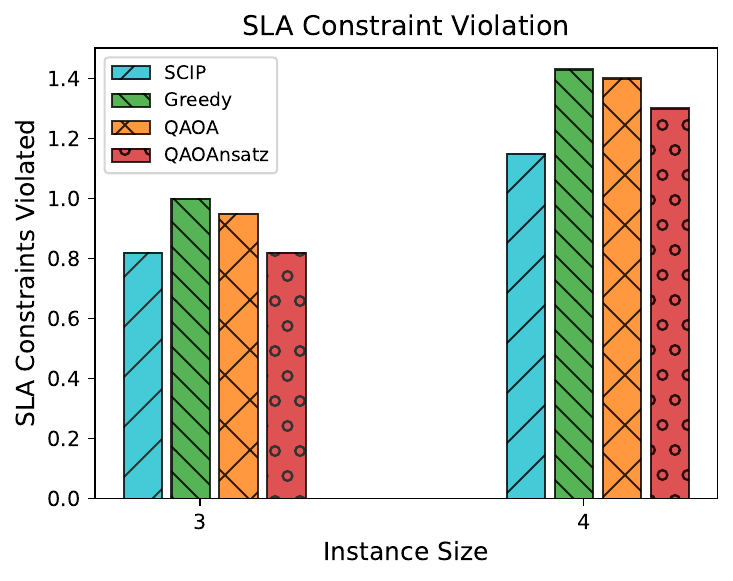}
    \caption{The SLA Constraint violations for small instance solvers.}
    \label{fig:small-pt}
\end{figure}

\begin{figure}
    \centering
    \includegraphics[width=\linewidth]{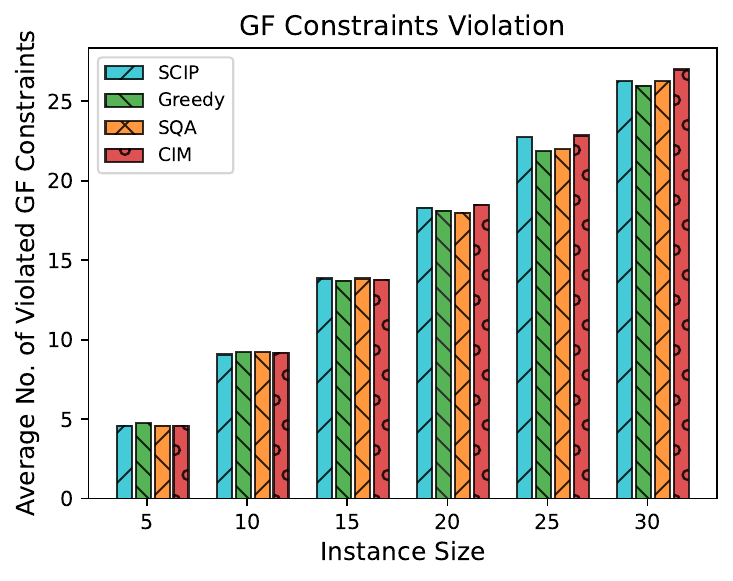}
    \caption{The GF Constraint violations for large instance solvers.}
    \label{fig:large-gf}
\end{figure}
\begin{figure}
    \centering
    \includegraphics[width=\linewidth]{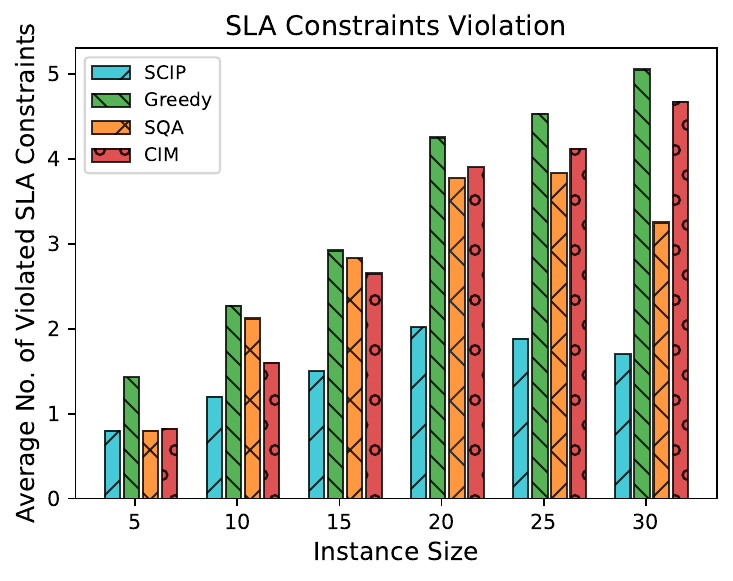}
    \caption{The SLA Constraint violations for large instance solvers.}
    \label{fig:large-pt}
\end{figure}

Constraint satisfaction behavior differs markedly between penalty-based and constraint-preserving approaches. For small instances~(Fig.\ref{fig:small-gf} and Fig.\ref{fig:small-pt}), both QAOA and QAOAnsatz exhibit violations of soft constraints, with QAOAnsatz consistently resulting in fewer violations compared to QAOA. This trend underscores the advantage of restricting the quantum evolution to the feasible subspace, which reduces the reliance on large penalty terms.

In the large-instance regime~(Fig.\ref{fig:large-gf} and Fig.\ref{fig:large-pt}), quantum-inspired solvers show higher rates of soft-constraint violations compared to classical solvers, particularly for geofencing constraints. While post-processing ensures satisfaction of hard constraints across all solvers, residual violations of soft constraints remain more prevalent in SQA and CIM solutions. These observations suggest a trade-off between solution quality, feasibility robustness, and computational cost across solvers.

\section{Conclusion}

In this work, we presented a comparative study of classical, quantum-inspired, and gate-based quantum solvers for the Rider–Order Assignment problem, a constrained optimization task motivated by real-world logistics operations. By formulating the problem as a constrained binary optimization model and evaluating solvers across appropriate instance-size regimes, we provided a systematic assessment of solver performance in terms of solution quality, runtime, and constraint satisfaction.

Our results indicate that classical solvers continue to offer superior performance in both runtime and solution quality for the problem sizes considered. The SQA solver demonstrates the ability to generate near-feasible solutions for larger instances, albeit at significantly higher computational cost. In comparison, the Coherent Ising Machine produces solutions of comparable quality to other quantum-inspired approaches while exhibiting lower runtime in the evaluated regimes. Gate-based variational quantum algorithms remain limited to very small instances, where constraint-preserving approaches such as QAOAnsatz show improved feasibility behavior compared to penalty-based QAOA.

This study highlights the practical strengths and limitations of emerging optimization paradigms when applied to realistic, constrained logistics problems. We hope that this work serves as a useful benchmark and reference point for future research on hybrid classical–quantum optimization methods in operational settings.

\textit{Code Availability:} The code used for the experiments is available on reasonable request.

\bibliographystyle{unsrt} % We choose the "plain" reference style
\bibliography{refs}

\end{document}